\documentclass[a4paper, 10pt]{article} 
\usepackage{graphicx, amssymb} 
\usepackage[sumlimits]{amsmath}
\usepackage{amsthm}
\usepackage{array}

\newtheorem{thm}{Theorem}

\newtheorem{prp}[thm]{Proposition}
\theoremstyle{remark}
\newtheorem{rem}[thm]{Remark}

\newcommand{\refsec}[1]{Section~\ref{sec:#1}}
\newcommand{\reffig}[1]{Figure~\ref{fig:#1}}
\newcommand{\refeqn}[1]{(\ref{eqn:#1})}
\newcommand{\reftbl}[1]{Table~\ref{tbl:#1}}
\newcommand{\refthm}[1]{Theorem~\ref{thm:#1}}
\newcommand{\refprp}[1]{Proposition~\ref{prp:#1}}
\newcommand{\refrem}[1]{Remark~\ref{rem:#1}}

\newcommand{\R}{\mathbb R}

\newcommand{\cD}{{\cal D}}

\newcommand{\cP}{{\cal P}}

\newcommand{\If}{I_{\mathrm{free}}}

\newcommand{\s}[1]{\left(#1\right)}

\DeclareMathOperator{\spn}{span}

\DeclareMathOperator{\wsize}{wsize}

\newcommand{\pfstart}{\begin{proof}} 
\newcommand{\pfend}{\end{proof}} 

\title{Span Programs for Functions with Constant-Sized 1-certificates}
\author{Aleksandrs Belovs\thanks{Faculty of Computing, University of Latvia, Raina bulv. 19, Riga, LV-1586, Latvia, stiboh@gmail.com.}}
\date{}

\begin{document}
\maketitle

\begin{abstract}
Besides the Hidden Subgroup Problem, the second large class of
quantum speed-ups is for functions with constant-sized 1-certificates.
This includes the OR function, solvable by the Grover algorithm,
the distinctness, the triangle and other problems. The usual way to solve them
is by quantum walk on the Johnson graph.

We propose a solution for the same problems using span programs. The
span program is a computational model equivalent to the quantum query
algorithm in its strength, and yet very different in its outfit.

We prove the power of our approach by designing a quantum algorithm for the
triangle problem with query complexity $O(n^{35/27})$ that is better than $O(n^{13/10})$
of the best previously known algorithm by Magniez {\em et al.}
\end{abstract}


\section{Introduction}
In this paper, we are interested in quantum query complexity of functions with 1-certificate complexity bounded by a constant. Research on quantum algorithms for such functions was launched shortly after the beginnings of quantum computation. The first example is that of Grover search~\cite{grover} for the OR function. The distinctness and the triangle problems also belong to this class.

One can distinguish two main design paradigms for quantum algorithms for such functions. The first one includes application of the Grover search and its close relative --- quantum amplitude amplification. This paradigm resulted in the algorithm for the collision problem  with complexity $O(n^{1/3})$ by Brassard {\em et al.} \cite{collision}, the $O(n^{3/4})$-algorithm for the distinctness problem by Buhrman {\em et al.} \cite{distinct1}, the $O(n^{10/7})$-algorithm for the triangle problem by Magniez {\em et al.} \cite{triangle}, and others.

The second paradigm is based on quantum walks on the Johnson graph. It was pioneered by Ambainis with his $O(n^{2/3})$-algorithm for the distinctness problem \cite{distinct2}. The triangle-finding algorithm with complexity $O(n^{13/10})$ by Magniez {\em et al.} \cite{triangle} also belongs to this class.

In this paper, we propose an approach to these problems using span programs. The span program is a computational model proven by Reichardt to be equivalent to the quantum query algorithm \cite{spanBig, spanOptimal}. Despite this equivalence, the actual applications of this model have been limited, mostly, to formulae evaluation \cite{formulae}.

We show that span programs are useful for other well-studied problems in quantum computation. We build analogues of the algorithms for the OR function and the distinctness problem with optimal complexity. We demonstrate the power of our approach by designing an algorithm for the triangle problem with complexity $O(n^{35/27})$, that is better than $O(n^{13/10})$ of the algorithm by Magniez {\em et al.}

The paper is organized as follows. In \refsec{prelim}, we review the notion of certificate complexity, define the problems being solved, and describe the span programs. In \refsec{learning}, we define learning graphs as our approach to functions with small 1-certificate complexity. In \refsec{stages}, we introduce the concept of stages that is illustrated by an example of a learning graph for the distinctness problem. In \refsec{symmetric}, we discuss how symmetry of the problem can be used in the design of learning graphs, and finally, in \refsec{triangle}, we give our algorithm for the triangle problem. 

\section{Preliminaries}
\label{sec:prelim}
\subsection{Quantum Query Algorithms for Functions with Constant-Sized 1-certificates}
For the basic concepts of quantum algorithms, a reader may refer to~\cite{chuang}. We are interested in query complexity of quantum algorithms, i. e., we measure the complexity of a problem by the number of queries to the input the best algorithm should make. Clearly, query complexity provides a lower bound on time complexity. For many algorithms, query complexity can be analysed easier than time complexity. For the definition of query complexity and its basic properties, as well as properties of certificate complexity, a good reference is~\cite{survey}.

Functions we work with in this paper have bounded 1-certificate complexity. Let us define what this means. Consider a multivariable function $f:[m]^n\to [2]$. By $[m]$, we denote the set $\{0,1,\dots,m-1\}$. We identify the set of input variables of $f$ with set $[n]$. An {\em assignment} is a function $\sigma: [n]\supset S\to [m]$. One should think of this function as fixing values for some input variables. We say input $x=\{x_i\}_{i\in [n]}$ {\em agrees} with assignment $\sigma$ if $\sigma(i) = x_i$ for all $i\in S$. The {\em size} of an assignment is the size of its domain $S$. 

Assignment $\sigma$ is called a {\em $b$-certificate} for $f$ if any input consistent with $\sigma$ is mapped to $b$ by $f$. The {\em certificate complexity} $C_x(f)$ of function $f$ on input $x$ is defined as the minimal size of a certificate for $f$ that agrees with $x$. The $b$-certificate complexity $C^{(b)}(f)$ is defined as $\max_{x\in f^{-1}(b)} C_x(f)$.

As it has been said, we are interested in algorithms for families of functions such that $C^{(1)}(f)$ remains bounded by a constant. Many quantum algorithms have been constructed for functions from this class. We work mostly with the following three functions.

\paragraph{OR function}
The simplest example of a function with constant 1-certificate complexity is the OR function. It is easy to see that $C^{(1)}(OR) = 1$, since it is enough to pick any input variable equal to 1. Note that 0-certificate complexity of this function is $n$.

The quantum algorithm for the OR function is one of the first quantum algorithms. It was invented by Grover \cite{grover} and it has query complexity $O(\sqrt{n})$. This is optimal \cite{groveroptimal}.

\paragraph{Distinctness Problem}
The {\em distinctness} function is function $f:[m]^n\to [2]$ such that $f(x_1,\dots,x_n)$ equals 1 iff there are equal elements among $\{x_1,\dots, x_n\}$. It has 1-certificate complexity $2$.

The first quantum algorithm for the distinctness problem had complexity $O(n^{3/4})$ and was due to Buhrman {\em et al.} \cite{distinct1}. This was later improved to $O(n^{2/3})$ by Ambainis \cite{distinct2}. This is the first natural problem solved by an algorithm based on a {\em quantum walk}. This algorithm is optimal, due to the result by Shi \cite{distinctOptimal}.

\paragraph{Triangle Problem}
Consider a full graph on $n$ vertices. The input variables of the function are in correspondence to the edges of the graph. Denote by $x_{ij}$ where $1\le i<j\le n$ the input variable corresponding to the edge joining vertices $i$ and $j$. The task is to detect whether there is a triangle with all edges marked by 1, i. e., whether there are indices $i<j<k$ such that $x_{ij} = x_{ik} = x_{jk} = 1$. Clearly, the 1-certificate complexity of the triangle problem is 3.

The best previously known quantum query algorithm for the triangle problem is due to Magniez {\em et al.} \cite{triangle} and has complexity $O(n^{13/10})$. We describe this algorithm in \refsec{triangle}. In the same section, we improve the complexity to $O(n^{35/27})$. The best known lower bound is just $\Omega(n)$.

\subsection{Span Programs}
\label{sec:span}
In this section, we define span programs following, mostly,~\cite{spanBig}. A span program $\cP$ is a way of computing a Boolean function $\{0,1\}^m\to\{0,1\}$. It is defined by
\begin{itemize}
\item A finite-dimensional inner product space $V=\R^n$. Reichardt {\em et al.} define span programs over $\mathbb{C}$, we find real span programs more convenient. Real span programs are known to be equivalent to the complex ones~\cite[Lemma 4.11]{spanBig};
\item A non-zero {\em target vector} $t\in V$;
\item A set of {\em input vectors} $I\subset V$. The set $I$ is split into the union of the set of {\em free input vectors} $\If$ and the collection of sets $\{I_{j,b}\}$ with $j=1,\dots,m$ and $b=0,1$: $I=\If\cup\bigcup_{j,b} I_{j,b}$. The input vectors of $I_{j,b}$ are {\em labeled} by the tuple of the $j$-th input variable $x_j$ and its possible value $b$.
\end{itemize}

For each input $x=(x_j)\in\{0,1\}^m$, define the set of {\em available} input vector as $I(x)=\If\cup\bigcup_{j=1}^m I_{j,x_j}$. Its complement $I\setminus I(x)$ is called the set of {\em false} input vectors. We say that $\cP$ evaluates to 1 on input $x$, iff $t\in\spn(I(x))$. In this way, span programs define total Boolean functions. One can define a span programs for a partial Boolean function as well, by ignoring the output of the program on the complement of the domain.

A useful notion of complexity for a span program is that of {\em witness size}. Assume, up to the end of the section, a span program $\cP$ calculates a partial Boolean function $f:\cD\to\{0,1\}$ with $\cD\subseteq \{0,1\}^m$. Let $A$ and $A(x)$ be matrices having $I$ and $I(x)$ as their columns, respectively.

If $\cP$ evaluates to 1 on input $x\in\cD$, a {\em witness} for this input is any vector $w\in \R^{|I(x)|}$ such that $A(x)w = t$. The size of $w$ is defined as its norm squared $\|w\|^2$. 

If, on contrary, $f(x)=0$ then a witness for this input is any vector $w'\in V$ such that $\langle w',t\rangle=1$ and that is orthogonal to all vectors from $I(x)$. Since $t\notin\spn(I(x))$, such a vector exists. The size of $w'$ is defined as $\|A^T w'\|^2$. Note that this equals the sum of squares of inner products of $w'$ with all false input vectors.

The witness size $\wsize(\cP, x)$ of span program $\cP$ on input $x$ is defined as the minimal size among all witnesses for $x$ in $\cP$. We also use notation
$$\wsize_b(\cP,\cD) = \max_{x\in \cD: f(x)=b} \wsize(\cP,x).$$
The witness size of $\cP$ is defined as
$$\wsize(\cP,\cD) = \sqrt{\wsize_0(\cP,\cD)\wsize_1(\cP,\cD)}.$$
This is not a standard definition, but it appears as equation (2.8) in~\cite{spanBig}.

The following important theorem is a combination of results from~\cite{spanOptimal} and~\cite{spanBig} and it shows why span programs are important for quantum computation:
\begin{thm}
\label{thm:span}
For any partial Boolean function $f\colon \{0,1\}^n\supset \cD\to \{0,1\}$ and for any span program $\cP$ computing $f$, there exists a 2-sided bounded error quantum algorithm calculating $f$ in $O(\wsize(\cP, \cD))$ queries.
\end{thm}

Thus, a search for a good quantum query algorithm is essentially equivalent to a search for a span program with small witness size.

\section{Learning graphs}
\label{sec:learning}
\subsection{Definitions}
\label{sec:def}
Our main model of computation for this paper is the {\em learning graph}, or just {\em L-graph}. It is a  directed acyclic connected graph with vertices being subsets of the set of input variables. Usually, we identify the latter with $[n]$, where $n$ is the number of input variables. Sometimes, we call the vertices of the learning graph {\em L-vertices}.

One may think of the learning graph as simulating the development of our knowledge on the input. Initially, we know nothing on the input, and it is represented by vertex $\emptyset$. When in vertex $S\subseteq [n]$, the values of the variables in $S$ have been learned. For any $j\in [n]\setminus S$, vertex $S$ can be connected to $S\cup \{j\}$ by an {\em arc}. This can be interpreted as querying the value of variable $x_j$. We say the arc {\em loads} element $j$. When talking about vertex $S$, we call $S$ the set of {\em loaded elements}.

Each arc $e$ is assigned a positive real number $w_e$ -- its {\em weight}. A learning graph is similar to a randomized decision tree with some differences. First of all, it is not a tree. Secondly, the values of the input variables do not figure in the model (see, however, \refrem{values}). And finally, there is no restriction on the weights of the arcs.

In order for a learning graph to calculate function $f$ correctly, the following property should be assured. For any $x\in f^{-1}(1)$, there exists a 1-certificate for $x$ contained in a vertex of the learning graph. We call such vertices {\em accepting}. For any correct learning graph, one can define its {\em complexity} as the geometrical mean of its {\em positive} and {\em negative complexities}. 

Let $E$ be the set of arcs. The negative complexity of the learning graph is defined as $\sum_{e\in E} w_e$. The positive complexity is more subtle. Fix an input $x\in f^{-1}(1)$ and consider a flow $p_e$ on the learning graph such that
\begin{itemize}
\item vertex $\emptyset$ is the only source of the flow, and it has intensity 1. In other words, the sum of $p_e$ over all $e$'s leaving $\emptyset$ is 1;
\item vertex $S$ is a sink iff it is accepting. That is, if $S\ne\emptyset$ and $S$ does not contain a 1-certificate for $x$ then, for vertex $S$, the sum of $p_e$ over all in-coming arcs equals the sum of $p_e$ over all out-going arcs.
\end{itemize}
The complexity of the flow is defined as $\sum_{e\in E} p_e^2/w_e$. The complexity for input $x$ is the minimum complexity over all possible flows satisfying these conditions. The positive complexity of the learning graph is the maximum complexity over all $x$'s such that $f(x)=1$.

We will also talk about flow $p_v$ through a vertex $v$. It is defined as the sum of $p_e$ over all arcs ending at $v$.

\begin{rem}
\label{rem:randomWalk}
Under a reasonable assumption $p_e\ge 0$, one can consider the above flow as a random walk. Indeed, consider the probability distribution on paths starting at $\emptyset$ and finishing in an accepting vertex, such that the probability an arc $e$ is used in the path is exactly $p_e$. In contrary to random walks used previously to build quantum walks, this is rather a random walk {\em through} the graph than {\em on} it. We utilize this probability language in \refsec{triangle}.
\end{rem}

For most of our applications, we consider only one certificate $\sigma$ for each input $x\in f^{-1}(1)$. We call the elements inside the domain of $\sigma$ {\em marked}. Then, the only task of the learning graph is to load all the marked elements. We construct learning graphs in order to minimize the complexity of this loading.

The following theorem links learning graphs and quantum query complexity.

\begin{thm}
\label{thm:learning}
For any learning graph for a function $f:[m]^n\to \{0,1\}$ with complexity $C$, there exits a bounded error quantum query algorithm for the same function with complexity $O(C\log m)$.
\end{thm}

The theorem is proven in \refsec{proof}, but before that we give a warm-up example.

\subsection{Grover Search}
We start with the description of a learning graph corresponding to the Grover algorithm. Recall, it calculates the OR function. An assignment $x_j\mapsto 1$ is a 1-certificate for every $j$.

The learning graph is quite simple. It has vertices $\emptyset$ and $\{1\}, \dots, \{n\}$. Vertex $\emptyset$ is connected by an arc of weight 1 to each of $\{i\}$'s. 

Clearly, the negative complexity is $n$. Let us calculate the positive complexity. Let $i$ be such that $x_i=1$. Define the flow equal to 1 on the arc from $\emptyset$ to ${\{i\}}$ and 0 for all other arcs. This gives the positive cost $1$. Hence, the complexity of the learning graph is $\sqrt{n}$ that coincides with the complexity of the Grover algorithm.

As well-known, if it is promised that there is either none of $x_i$'s equal to 1, or at least $r$ of them, the complexity of the Grover algorithm becomes $O(\sqrt{n/r})$. This can be shown using the same learning graph. Let $M$ be the set of input variables equal to 1. Define the flow as $1/|M|$ along an arc to a vertex containing an element of $M$, and 0, otherwise. This gives the positive complexity $|M|\frac{1}{|M|^2}=1/|M|$. Hence, the total complexity is $\sqrt{n/r}$.

This illustrates the main point about positive complexity. We want to distribute the flow as evenly as possible along as many paths as possible. Doing so reduces the complexity because of convexity of the square function.

\subsection{Proof of \refthm{learning}}
\label{sec:proof}
We aim to apply \refthm{span}, i. e., to build a span program and estimate its witness size. Let us start with the Boolean case $m=2$.

\paragraph{Description of the Span Program}
Let us describe the vector space of the span program. Each vertex $S$ of the learning graph is represented by $2^{|S|}$ vectors $\{t_\sigma\}$ where $\sigma$ is an element of $[2]^S$. We assume all this vectors are orthonormal. One may think of $t_\sigma$ as representing the values learned while querying elements of $S$, while vertex $S$ represents the sole fact the variables have been queried. Vector $t_\emptyset$, that corresponds to vertex $\emptyset$, is the target of the span program. If $\sigma:S\to [2]$ is a 1-certificate for $f$, $t_\sigma$ is a free input vector.

Consider an arc $e$ from $S$ to $S\cup\{j\}$ with weight $w_e$. For each vector $t_\sigma$ such that $\sigma$ has domain $S$, we add two input vectors
\begin{equation}
\label{eqn:vektora}
\sqrt{w_e} (t_\sigma - t_{\sigma \cup \{j\mapsto b\}} ), \qquad b=0,1.
\end{equation}
Here $\sigma \cup\{j\mapsto b\}$ is the assignment with domain $S\cup\{j\}$ that maps $i$ to $\sigma(i)$ for $i\in S$ and maps $j$ to $b$. Each of these two vectors is labeled by value $b$ of variable $x_j$.

\paragraph{Negative Witness Size}
Let us describe the negative witness $w'$ of the span program. Fix an input $x\in f^{-1}(0)$. For each $S$, define $\iota(S)$ as the only assignment $S\to[2]$ agreeing with the input. For each $t_\sigma$, we define $\langle w', t_\sigma\rangle = 1$ if $\sigma$ agrees with the input. Otherwise, we define $\langle w', t_\sigma\rangle = 0$.

Consider a free input vector of the form $t_\sigma$. Since $f(x)=0$, and $\sigma$ is a 1-certificate, $\sigma$ does not agree with the input. By the construction, $t_\sigma$ is orthogonal to the witness.

Consider an available input vector of the form \refeqn{vektora}. There are two cases
\begin{enumerate}
	\item Inner product $\langle w', t_\sigma\rangle$ equals 0. In this case $\sigma$ does not agree with the input, and, {\em a fortiori}, none of $\sigma\cup\{j\mapsto 0\}$ and $\sigma\cup\{j\mapsto 1\}$ agrees with the input. Hence, both vectors of \refeqn{vektora} are orthogonal to the witness.
	\item Inner product $\langle w', t_\sigma\rangle$ equals 1. In this case $\langle w', t_{\sigma\cup\{j\mapsto b\}}\rangle = 1$ if $x_j=b$.
\end{enumerate}
In both cases, the available input vector of \refeqn{vektora} is orthogonal to the witness. This proves that $w'$ indeed is a negative witness.

Let us calculate the size of $w'$. Let $e$ be an arc from $S$ to ${S\cup\{j\}}$. We claim there is exactly one input vector that arises from $e$ and isn't orthogonal to the witness. Let $\sigma$ have domain $S$. By the first point above, if $\sigma$ does not agree with the input, both input vectors of \refeqn{vektora} are orthogonal to $w'$. If $\sigma=\iota(S)$, the inner product of the false input vector from \refeqn{vektora} and $w'$ is $\sqrt{w_e}$. By summation over all arcs, we have the size of $w'$ equal to $\sum_e w_e$, i. e., to the negative complexity of the learning graph.

\paragraph{Positive Witness Size}
Now, let us calculate the positive witness size. Fix an input $x$ such that $f(x)=1$, and let $p_e$ be the corresponding flow. We will give a linear combination of the available input vectors that equals $t_\emptyset$.

Let $e$ be an arc from $S$ to ${S\cup\{j\}}$ with weight $w_e$. Let $\sigma=\iota(S)$ and take the available input vector from \refeqn{vektora} with coefficient $p_e/\sqrt{w_e}$. Multiplied by the coefficient, the vector equals $p_e(t_{\iota(S)} - t_{\iota(S\cup\{j\})})$.

Suppose vector $S$ be a sink. Then, $t_{\iota(S)}$ is a free input vector. Take it with the coefficient equal to the difference of the in-flow to $S$ and the out-flow of $S$. 

By the properties of the flow, the sum of all these vectors equals $t_\emptyset$ that is the target vector. The witness size is $\sum_e p_e^2/w_e$, i. e., the positive complexity of the learning graph. This proves the theorem for the Boolean case.

\paragraph{Non-Boolean Case}
Now consider the case $m>2$. Fix a representation of elements of $[m]$ using $k=\lceil\log m\rceil$ bits. For $j\in [n]$, construct a set $B_j$ of $k$ Boolean variables representing $x_j$. Consider Boolean function $f': [2]^{\cup_j B_j}\to [2]$ obtained from $f$ by encoding the inputs variables. One can construct a learning graph $G'$ for $f'$ from the learning graph $G$ for $f$ in the following way. Replace each vertex $S$ by $S'=\cup_{j\in S} B_j$. For an arc from $S$ to $S\cup\{j\}$ with weight $w_e$, fix an arbitrary order $y_1,\dots,y_k$ of elements in $B_j$ and represent the arc as path $S', S'\cup\{y_1\},\dots, S'\cup\{y_1,\dots,y_k\}$ in $G'$ of $k$ arcs, each of weight $w_e$.

Clearly, the negative complexity of $G'$ is $k$ times the negative complexity of $G$. Accepting vertices of $G$ are transformed into accepting vertices of $G'$, and each flow through $G$ can be transformed into a flow through $G'$ in an obvious way. This increases the complexity of the flow $k$ times. Hence, the complexity of $G'$ is at most $k$ times the complexity of $G$, and, for $G'$, we can apply the construction from the first part of the proof.

\subsection{Additional Remarks}
The construction used in the proof of the non-Boolean case of \refthm{learning} is a special case of multiplexor from \cite{rank}. The main reason behind the appearance of the $\log m$ factor in the witness size of the span program is the representation of an $m$-ary variable by a set of $\log m$ Boolean variables. It is tempting to claim that this factor can be removed, if one allows queries to the $m$-ary variable directly, as it usually is done in quantum query algorithms for non-Boolean functions. Unfortunately, we do not know yet how to use such queries in span programs.

Often, It will be convenient to use more than one vertex with the same subset $S$ in the learning graph. We will distinguish them using some additional labels. \refthm{learning} also holds for such learning graphs. One may prove that by noticing that the proof of the theorem does not change if one adds additional labels to the vertices. Another, more illuminating reasoning is as follows. 

Assume we have a vertex $S$ in a learning graph $G$ with arcs $e_1,\dots, e_k$ going to vertices $(S',1), \dots, (S',k)$, respectively, that represent the same subset of input variables $S' = S\cup\{j\}$. Let the weights of the arcs be $w_1,\dots,w_k$. Replace the $k$ vertices by one vertex $S'$ and $k$ arcs by one arc $e$ connecting $S$ to $S'$ of weight $w_1+\cdots+w_k$. Clearly, this does not change the negative complexity of the learning graph.

For the positive complexity, assume we have a flow on $G$ with flow through $e_i$ equal to $p_i$. We can construct the corresponding flow on $G'$ by sending flow $p_1+\dots+p_k$ through $e$. The positive complexity can decrease only, because
\[\frac{(p_1+p_2+\cdots+p_k)^2}{w_1+\cdots+w_k} \le \frac{p_1^2}{w_1}+\cdots+\frac{p_k^2}{w_k}.\]
The last inequality follows from the Jensen's inequality for the square function
\[(\alpha_1 x_1+\cdots + \alpha_k x_k)^2 \le \alpha_1 x_1^2 + \cdots+ \alpha_k x_k^2,\]
with $\alpha_i = w_i/(w_1+\cdots+w_k)$ and $x_i = p_i/\alpha_i$.

One can transform a learning graph with additional labels into a learning graph as defined in \refsec{def} using the above transformation repeatedly, and the complexity can decrease only. Hence, the definition of the learning graph from \refsec{def} is optimal from the point of view of complexity.

\begin{rem}
\label{rem:values}
In our construction of the learning graph, the weights of the arcs leaving vertex $t_S$ do not depend on the value of the variables inside $S$. By analysing the proof of \refthm{learning}, mostly the fact vertex $S$ is being split into vectors $\{t_\sigma\}$, one can see that it goes through in more general settings. Namely, one can define weights of the arcs leaving vertex $S$ in dependence on the values of the variables inside $S$. When calculating the positive and the negative complexities, one sums up only the arcs that match the input. In this case, the negative complexity also depends on the input. We didn't find this model useful for our applications, but it is possible that for other problems it could provide some improvements.
\end{rem}

\section{Distinctness Problem: Learning by Stages}
\label{sec:stages}
In this section, we describe our approach to the learning graphs using {\em stages}. We illustrate our construction with the example of the distinctness problem.

Stage is a slice of the learning graph consisting of arcs with similar functionality. Functionality is described with respect to the flows used in the definition of the positive complexity. For each input $x\in f^{-1}(1)$, we select a 1-certificate $\sigma$. Recall that the elements of the domain of $\sigma$ are called marked. E. g., for the distinctness problem, we mark any two elements $a$ and $b$ having the same value.

Different stages represent different relation of the arcs used in the flow to the marked elements. For each problem, the number of stages doesn't depend on the size of the instance, i. e., it equals $O(1)$. For the distinctness problem, for example, we have three stages as in \reftbl{distinctness}, independently on $n$. Thus, on stage I non-zero flow have only arcs not loading $a$ or $b$, on stage II --- arcs loading $a$, on stage III --- $b$. The stages are written in the order they are used in the flow. I. e., any random walk through the learning graph will use arcs of stage I first, then arcs of stage II and so on.

\begin{table}[htb]
\begin{tabular}{rp{13cm}}
\hline
I.& Load $r-2$ items different from $a$ and $b$.\\
II.& Load $a$.\\
III.& Load $b$.\\
\hline
\end{tabular}
\caption{Stages for the distinctness problem.}
\label{tbl:distinctness}
\end{table}

Any stage performs a transition from a subset of L-vertices to another subset of L-vertices. If there are $k$ stages, there are $k+1$ subsets: denote them by $V_0,\dots,V_k$: the $i$-th stage moves from $V_{i-1}$ to $V_i$. The subset $V_0$ consists of the initial vertex $\emptyset$.

Now we describe the stages in more detail. Consider the $i$-th stage. Vertex $S\in V_{i-1}$ may be connected by a {\em transition} $e$ to vertex $S'\in V_i$ only if $S\subset S'$. The {\em length} $\ell(e)$ of the transition is defined as $|S'\setminus S|$. We denote the set of transitions of stage $i$ by $E_i$. The set of all transitions is denoted by $E$.

We define a {\em reduced learning graph} as a graph on the vertex set $V_0\cup V_1\cup\cdots \cup V_k$ and transitions instead of arcs. For an input $x\in f^{-1}(1)$, one can define a flow through the reduced graph in the same way as for a learning graph. We select a flow $p_e(x)$ on the reduced graph for each input $x\in f^{-1}(1)$. We break the complexity of the whole learning graph into complexities of the individual stages. A complexity of stage $i$ is defined by
\[C_i = \max_{x\in f^{-1}(1)} \sqrt{\s{\sum_{e\in E_i} \ell(e) w_e} \s{\sum_{e\in E_i} \frac{\ell(e) p_e(x)^2}{w_e}} }. \]

\begin{prp}
\label{prp:summa}
For a reduced learning graph $G$ with $k$ stages and complexities $C_i$ of each stage, one can build a learning graph of complexity $O(C_1+\cdots+C_k)$.
\end{prp}

\pfstart
At first, we transform the reduced learning graph into a learning graph. Let $e$ be a transition between $S$ and $S'$ with weight $w_e$. Fix an arbitrary ordering of elements of $S'\setminus S = \{s_1,\dots,s_{\ell(e)}\}$. Represent $e$ by path $S, S\cup\{s_1\}, S\cup\{s_1,s_2\},\dots, S'\setminus\{s_{\ell(e)}\}, S'$ in the learning graph. In order to simplify calculations, we assume paths for different transitions do not intersect. In other words, we make a unique copy of a vertex for each transition that uses it as an internal vertex. We assign weight $w_e$ to each arc of the path. Similarly, if $p_e$ is a flow through the transition, we set the flow through each arc of the path equal to $p_e$.

It is easy to see that the negative and the positive complexities, when we consider only arcs on stage $i$, are, respectively,
\[N_i = \sum_{e\in E_i} \ell(e) w_e\qquad\text{and}\qquad P_i = \sum_{e\in E_i} \frac{\ell(e) p_e^2}{w_e}.\]

Now divide the weights of arcs on stage $i$ by $N_i$. Clearly, the negative complexity of the whole learning graph becomes $k=O(1)$. The complexity of the learning graph becomes 
\[O\s{\sqrt{\sum_{i=1}^k N_i P_i}}  = O\s{\sqrt{\sum_i C_i^2}} = O\s{\sum_i C_i}.\qedhere \]
\pfend

We call a transition {\em valid}, if it satisfies the condition stated in the description of the stages, i.e., if it is used in the flow. Whether a transition is valid depends on the input $x$, more precisely, on the set of marked elements. For the distinctness problem, a transition on stage II is valid if it loads $a$ and originates in a vertex without $b$. A transition on stage III is valid if it loads element $b$ and comes from a vertex containing $a$. Similarly, we call a vertex {\em valid} if it has non-zero flow through it.

\begin{figure}[tbh] 
\centering \includegraphics[width=9cm]{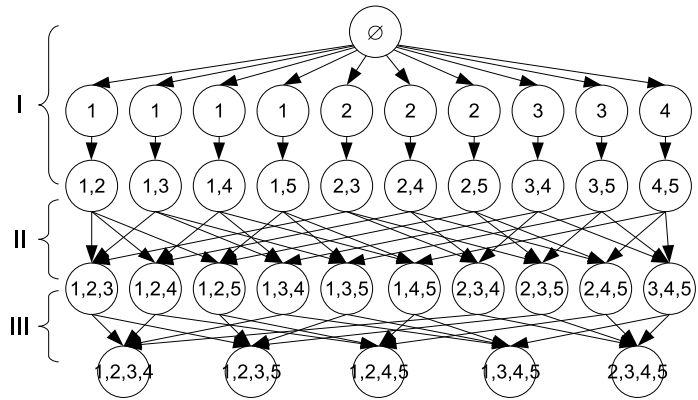} 
\caption{The learning graph for the distinctness problem in case $n=5$ and $r=4$. Stages I, II and III shown.}
\label{fig:distinct1}
\end{figure}

At the moment of construction, we do not know which transitions are valid, so we assume all possibilities. E. g., for the distinctness problem, subsets $V_1,V_2$ and $V_3$, are vertices of the learning graph with $r-2$, $r-1$ and $r$ elements loaded, respectively, and we add transitions everywhere possible. For an illustration refer to \reffig{distinct1}.

Let us describe the flow for the distinctness problem. It is already implicitly described in the description of the stages. On stage I, we let a flow ${n-2\choose r-2}^{-1}$ along any transition ending in a vertex with both $a$ and $b$ not loaded. Denote such a vertex by $v$. Then we forward the flow along the transition to $v\cup\{a\}$ and then to $v\cup\{a,b\}$ on stages II and III, respectively. The last L-vertex is a sink, so we may stop. Let us calculate the exact expressions for complexities of the stages of the learning graph for the distinctness problem. We assume all transitions have weight 1. We have
\begin{align}
\label{eqn:C1}
C_1 &= \sqrt{\s{{n\choose r-2}(r-2)} \s{{n-2\choose r-2}(r-2){n-2\choose r-2}^{-2}} } = O(r), \\
\label{eqn:C2}
C_2 &= \sqrt{\s{(n-r+2){n\choose r-2}} \s{{n-2\choose r-2}{n-2\choose r-2}^{-2}} } = O(\sqrt{n}), \\
\label{eqn:C3}
C_3 &= \sqrt{\s{(n-r+1){n\choose r-1}} \s{{n-2\choose r-2}{n-2\choose r-2}^{-2}} } = O(n/\sqrt{r}).
\end{align}
Hence, the total complexity, by \refprp{summa}, is
\[O(r + \sqrt{n} + n/\sqrt{r}), \]
that attains its optimal value $O(n^{2/3})$ when $r=n^{2/3}$.

\section{Using symmetry}
\label{sec:symmetric}
In the following, we study symmetric problems, i. e., problems that stay invariant under a wide group of permutations. For instance, the OR problem and the distinctness problem stay invariant under the action of the full symmetric group on the input variables. The triangle problem is invariant under permuting vertices of the graph. Denote the group that leaves the problem invariant by $\Sigma$.

In fact, the symmetry of the problem is not the main thing that concerns us. What we really use is that it is possible to select a collection of sets of marked elements so that, firstly, they cover all possible inputs in $f^{-1}(1)$ and, secondly, group $\Sigma$ acts transitively on this collection. Because of the transitivity, there is no real difference in the possible sets of marked elements. This property also can be fulfilled for non-symmetric functions. For example, if all 1-certificates of a function $f$ have size at most $k$, one can consider all $k$-subsets as possible sets of marked elements. But this may result in a sub-optimal algorithm.

\subsection{Transitions}
We assume the learning graph stays invariant under the action of $\Sigma$. Call two transitions of the reduced learning graph {\em equivalent}, if one can be transformed into another by an element of $\Sigma$. I. e., if transition $e$ is from $S$ to $T$ and $e'$ is from $S'$ to $T'$, they are equivalent if there exists $\sigma\in\Sigma$ such that $\sigma(S)=S'$ and $\sigma(T) = T'$. We assume equivalent transitions have equal weights.

We define {\em speciality} $\tau(e)$ of a transition $e$ as the ratio of the size of the equivalence class containing $e$ to the number of valid transitions in it. It equals the inverse of the probability of obtaining a valid transition when a random permutation from $\Sigma$ is applied to $e$. 

Assume a flow $p_e(x)$ has been fixed for each input $x\in f^{-1}(1)$. We say the learning graph is {\em symmetric} on stage $i$ if the following two conditions hold:
\begin{itemize}
\item speciality of each equivalence class does not depend on the input $x$;
\item flows through all valid transitions in an equivalence class are equal, and its common value does not depend on the input, but only on the equivalence class.
\end{itemize}

Similarly, one can define equivalence of vertices in set $V_i$ and define speciality of a vertex and symmetry of the learning graph on set $V_i$.

Define the {\em average length} $L_i$ of stage $i$ by $\sum_{e\in E_i} p_e\ell(e)$. If the flow is symmetric, this quantity does not depend on the input. Let $T_i$ denote the maximal speciality of a transition on stage $i$. 

\begin{thm}
\label{thm:symmetric}
Assume the flow is symmetric on stage $i$. Then, the complexity of the stage is at most $L_i\sqrt{T_i}$.
\end{thm}

\pfstart
Consider $e\in E_i$ and let $p'(e)$ be the flow on a valid transition equivalent to $e$. By the above assumptions, $p'(e)$ does not depend on the input. Hence, it is possible to define the weight of transition $e$ equal to $p'(e)$. Then the complexity of the stage is 
\[ \sqrt{\s{\sum_{e\in E_i} \ell(e) p'(e) } \s{\sum_{e\in E_i} \frac{\ell(e) p_e^2}{p'(e)}}} = 
\sqrt{\s{\sum_{e\in E_i} \ell(e)\tau(e) p_e } \s{\sum_{e\in E_i} \ell(e) p_e}} \le L_i\sqrt{T_i}. \qedhere
\]
\pfend

We leave as an exercise for the reader to check that the flow for the distinctness problem in \reftbl{distinctness} satisfies conditions of \refthm{symmetric} and to check \reftbl{distinctnessParam}, where each entry is given up to a constant factor. Specialities are calculated in \refeqn{C1}---\refeqn{C3}.
\begin{table}
\centering 
\begin{tabular}{|r|ccc|}
\hline
Stage & I & II & III \\
\hline
Speciality & 1 & $n$ & $n^2/r$ \\
Length & $r$ & 1 & 1 \\
\hline
\end{tabular}\caption{Parameters (up to a constant factor) of the stages of the learning graph for the distinctness problem of \reftbl{distinctness}.}
\label{tbl:distinctnessParam}
\end{table}
Thus, the total complexity of the learning graph for the distinctness problem is $r + \sqrt{n} + n/\sqrt{r}$ that is optimized when $r=n^{2/3}$.

\subsection{Subroutines}
One can use subroutines in learning graphs in the following manner. Suppose we have a vertex $S\subseteq [n]$ in a learning graph $G$. One can treat $S$ as the initial vertex for a problem with variable set $[n]\setminus S$. Let $G'$ be a learning graph for this new problem. One can append $G'$ after $S$ in $G$.

Suppose, for some input $x\in f^{-1}(1)$, vertex $S$ has in-flow $\delta$. Let $M$ be the set of marked elements. One can take a flow in $G'$ that corresponds to the set of marked elements $M\setminus S$. As in an independent subroutine, the out-flow of vertex $S$ is 1. Let $P$ be the complexity of the flow. If one multiplies the flow through each arc of $G'$ by $\delta$, this makes $S$ have the same in- and out-flows, and the contribution towards the complexity of flow on $G$ from all arcs inside $G'$ becomes $\delta^2P$.

We allow a {\em subroutine stage} in the reduced learning graph. It is the last stage, and it starts in vertex set $V_k$. For each vertex of this set, we apply the procedure described in the previous paragraph. For simplicity, we assume there is exactly one subroutine for each vertex.

One can also consider a subroutine on a strict subset $I\subset [n]\setminus S$ of the set of remaining input variables. In this case, for each valid vertex of $t_S\in V_k$, subset $I\cup S$ should contain all marked elements.

One can apply symmetry for the subroutine stage, similarly as it is done for transitions. Let $\ell(v)$ be the complexity of the subroutine appended to vertex $v$. Define the {\em average complexity} $L$ of the subroutine stage as $\sum_{v\in V_k} p_v \ell(v)$. Let $T$ denote the maximal speciality of a vertex in $V_k$.
\begin{thm}
\label{thm:subroutine}
Suppose the flow is symmetric for vertex set $V_k$. Then, the complexity of the subroutine stage is $L\sqrt{T}$.
\end{thm}

\pfstart
The proof is similar to that of \refthm{symmetric}. Consider $v\in V_k$ and let $p'(v)$ be the flow through a valid vertex equivalent to $v$. Assume the negative and the positive complexities of the subroutine after $v$ both are $\ell(v)$. Multiply the weights of all arc of the subroutine by $p'(v)$. Then the complexity of the subroutine stage is 
\[ \sqrt{\s{\sum_{v\in V_k} \ell(v)p'(v)} \s{\sum_{v\in V_k} \frac{\ell(v)p_v^2}{p'(v)}}} = 
\sqrt{\s{\sum_{v\in V_k} \ell(v)\tau(v) p_v} \s{\sum_{v\in V_k} \ell(v)p_v}} \le L\sqrt{T}.\qedhere \]
\pfend

\section{Triangle Problem}
\label{sec:triangle}
Let us start with a learning graph for the triangle problem corresponding to the algorithm from \cite{triangle}. Denote the vertices of the triangle by $a$, $b$ and $c$. Stages of the learning graph are described in \reftbl{triangleOld}.

\begin{table}[htb]
\begin{tabular}{rp{13cm}}
\hline
I.& Load a complete subgraph on $r-2$ vertices that does not contain vertices $a,b$ and $c$.\\
II.& Load all edges connecting $a$ to the subgraph.\\
III.& Load all edges connecting $b$ to the subgraph (including $a$). Thus result in a complete subgraph on $r$ vertices with $a$ and $b$ inside, but $c$ outside.\\
IV.& Load $\ell$ edges that connect $c$ to $\ell$ vertices of the subgraph other than $a$ and $b$.\\
V.& Load edge $\widetilde{ac}$.\\
VI.& Load edge $\widetilde{bc}$.\\
\hline
\end{tabular}
\caption{Stages for the triangle problem according to the algorithm of \cite{triangle}.}
\label{tbl:triangleOld}
\end{table}

\begin{table}
\centering 
\begin{tabular}{|r|cccccc|}
\hline
Stage & I & II & III & IV & V & VI\\
\hline
Speciality & 1 & $n$ & $n^2/r$ & $n^3/r^2$ & $n^3/r$ & $n^3/\ell$ \\
Length & $r^2$ & $r$ & $r$ & $\ell$ & 1 & 1\\
\hline
\end{tabular}
\caption{Parameters (up to a constant factor) of the stages of the learning graph of \reftbl{triangleOld}}
\label{tbl:triangleOldParam}
\end{table}

It is not hard to check that the obvious flow on this L-graph satisfies conditions of \refthm{symmetric}. The parameters of the stages are in \reftbl{triangleOldParam}. Hence, by \refthm{symmetric}, the complexity of the L-graph is
\[O(r^2 + r\sqrt{n} + n\sqrt{r} + n^{3/2}\ell/r + n^{3/2}/\sqrt{r} + n^{3/2}/\sqrt{\ell}).\]
It is optimized when $r=n^{3/5}$ and $\ell = n^{2/5}$. The optimum is $O(n^{13/10})$. 

Let us analyze \reftbl{triangleOldParam}. The key point is to minimize the speciality of stage VI using previous stages. Also, the complexities of stages II and V are majorized by the complexities of stages III and VI, respectively. Thus, we should concentrate on stages I, III, IV and VI. Their complexities are $n^{12/10}$, $n^{13/10}$, $n^{13/10}$ and $n^{13/10}$, respectively. It looks like resources of stage I are not used to the full extent. The problem is in stage III: it is too long. The main idea behind our construction is to reduce the length of stage III.

\begin{thm}
\label{thm:triangle}
The triangle problem can be solved in $O(n^{35/27})$ quantum queries with a bounded error.
\end{thm}

Note that $35/27 \approx 1.2963 < 1.3$ of the previous algorithm.

\pfstart
First, let us agree that for the vertices of the learning graph we will use term L-vertex in order to distinguish them from the vertices of the input graph. In the L-graph we will have transitions, whereas in the input graph we have edges.

We use a symmetric approach based on stages as in \refsec{symmetric}. The symmetry group consists of all permutations of vertices. Unlike the learning graph for the distinctness problem, where all valid transitions on the same stage had the same flow through them, this time the flow will vary inside one stage. Thus, when defining stages, we will not only specify the relation of the transitions used in the flow to the marked elements, but also the value of the flow going through them.

To define the flow, we use the language of probability from \refrem{randomWalk}. Before we start the description of the stages, let us clarify our usage of term ``random subgraph on $k$ vertices''. By this we understand a subgraph of the complete graph on $n$ vertices constructed in the following way. Take a subset $U$ of $k$ vertices uniformly at random. Then, for each edge with both ends in $U$, add it to the subgraph with some prescribed probability $s$, independently at random. The description of the random subgraph contains both the vertex set and the selected edges. It is important to add the vertex set to the description, as it might happen some of the vertices inside $U$ end up with degree 0.

\begin{table}[htb]
\begin{tabular}{rp{13cm}}
\hline
I.& Load a random subgraph on $r-2$ vertices that does not contain vertices $a,b$ and $c$.\\
II.& Randomly load edges connecting $a$ to the vertices of the subgraph.\\
III.& Randomly load edges connecting $b$ to the subgraph (including $a$). The result is a random subgraph on $r$ vertices with $a$ and $b$ inside the vertex set, but not $c$.\\
IV.& Select those L-vertices that do not contain edge $\widetilde{ab}$ and contain at least $sr^2/4$ edges.\\
V.& Add edge $\widetilde{ab}$.\\
VI.& Add vertex $c$.\\
VII.& Use a subroutine from \reftbl{distinctness} to load edges $\widetilde{ac}$ and $\widetilde{bc}$ out of all edges connecting $c$ to the vertices of the subgraph.\\
\hline
\end{tabular}
\caption{Stages for the triangle problem.}
\label{tbl:triangle}
\end{table}

The stages of the algorithm are described in \reftbl{triangle}. We will describe each stage in more detail later, but let us say now that stage IV is different from the others. It is not a stage in the previous definition, it is a modifier for the flow before it. The reason behind its inclusion is that we want to apply \refthm{symmetric} to all stages, although the actual flow does not satisfy the conditions of the theorem. We solve this problem by constructing an ideal flow that satisfies conditions of \refthm{symmetric}, and prove that the actual flow have at most constant times larger complexity than the ideal one.

It is straightforward to check that conditions of \refthm{symmetric} are satisfied for stages I to III. For example, the flow through a valid transition on stage III that originates in a subgraph having $m$ edges and loads $k$ edges is
\[ {n-3\choose r-2}^{-1} s^{m+k} (1-s)^{{r\choose 2} - m - k}. \]

The speciality of transitions on stages I, II and III are $O(1)$, $O(n)$ and $O(n^2/r)$, respectively, independently on the equivalence class of the transition by the same argument as for the distinctness problem. The average length of a transition on stage I is $O(sr^2)$ by the standard probabilistic argument. Similarly, conditioned on leaving a fixed valid L-vertex, the average length of a transition on stages II and III is $O(sr)$. Hence, it coincides with the (unconditioned) average length of a transition on stages II and III.

Now, we switch to stage IV. The set of L-vertices before and after the stage is the same and the stage uses no transitions. What it does, it modifies the flow on stages I to III constructed so far. Consider L-vertices after stage III. Additionally to the condition to contain $a$ and $b$ and to not contain $c$, that is stated for state III, we require the subgraph to not contain edge $\widetilde{ab}$ and to have at least $sr^2/4$ edges. 

Denote the probability of the L-vertex of the constructed random walk to satisfy the new constraints after stage III by $p$. By an easy probabilistic argument, under reasonable assumptions ($s=o(1)$ and $sr^2=\omega(1)$), the probability is $1-o(1)$. Assume the instance is large enough, so that $p\ge 1/2$. Then, we scale up the flow $1/p$ times, and remove all flow going to the bad L-vertices. After that the intensity of the flow is 1 again, and the flow on each transition has increased at most 2 times. Hence, this operation changes the complexity of stages I-III at most by a factor of 4, hence, we can ignore it in our calculations.

Now, consider stage V. Before the filtering on stage IV, the flow through a valid L-vertex depended on the number of edges in it only. Hence, after the filtering, the flow also depends on the number of edges only. Moreover, we can remove L-vertices with less than $sr^2/4$ edges in them, because they never have flow through them. Denote a flow through a valid L-vertex with $m$ edges by $p_m$. On stage V, we connect an L-vertex with all possible L-vertices where an edge connecting two vertices of the subgraph is added.

Let us calculate the speciality of a transition on stage V. For each equivalence class, the probability a random permutation of vertices identifies the edge being added with $\widetilde{ab}$ is exactly $\frac{2}{n(n-1)}$. Moreover, provided that this happens, the probability that $c$ is not used in the vertex set of the subgraph is $(n-r)/(n-2)$. Hence, under assumption $r<n/2$, the speciality is $O(n^2)$. The length of the stage is, clearly, 1.

Stage VI does not add any edges, hence its complexity is 0. It adds a special vertex to the description of a L-vertex. The special vertex lies outside the vertex set of the subgraph. In this way, it increases the speciality of a L-vertex. Let us calculate it. If a random permutation is applied, the probability the special vertex gets mapped to $c$ and both $a$ and $b$ are in the vertex set of the subgraph is $O(r^2/n^3)$. This should be multiplied by the probability edge $\widetilde{ab}$ is among the edges of the subgraph for a fixed choice of the vertex set containing both $a$ and $b$. This probability is $O(m/r^2)$. Hence, the speciality of the L-vertex is $O(n^3/m) = O(n^3/sr^2)$ because of the filtering on stage IV. Since, as proven in \refsec{symmetric}, the complexity of the distinctness-type learning graph is $O(r^{2/3})$, the complexity of the last stage is $O(n^{3/2}s^{-1/2}r^{-1/3})$ by \refthm{subroutine}.

\begin{table}
\centering 
\begin{tabular}{|r|ccccc|}
\hline
Stage & I & II & III & V & VII\\
\hline
Speciality & 					1 & 			$n$ & 	$n^2/r$ & $n^2$ &	$n^3/sr^2$  \\
Length/Complexity & 	$sr^2$ & 	$sr$ & 	$sr$ & 		$1$ &  	$r^{2/3}$\\
\hline
\end{tabular}
\caption{Parameters (up to a constant factor) of the stages of the learning graph of \reftbl{triangle}}
\label{tbl:triangleParam}
\end{table}

Our estimates are summarized in \reftbl{triangleParam}. Adding everything up, the complexity of the learning graph is
\[O(sr^2 + sr\sqrt{n} + sn\sqrt{r} + n + n^{3/2}s^{-1/2}r^{-1/3}).\]
This is optimized when $r=n^{2/3}$ and $s = n^{-1/27}$. The optimum equals $O(n^{35/27})$. Note that complexities of stages I, III and VII are equal this time.
\pfend

\section{Summary and Future Work}
An approach towards quantum algorithms for functions with small 1-certificate complexity using span programs has been proposed in the paper. It seems to be at least as powerful as the previous approach of quantum walk on the Johnson graph.

The analysis of the algorithm contains no spectral analysis. It uses an optimization of a quadratic function over the set of flows.

A nice property of the new approach is that is has build-in tools for amortization. If some computational paths in the program take less queries than the others, the complexity of the program is calculated using the average rather than the maximal cost, as it is in the previous approaches.

Disadvantages of our method include that it only can be used for the query complexity, and the additional $\log m$ factor for non-Boolean functions.

The future work could consist in designing new algorithms based on this framework. Another interesting problem is to remove the $\log m$ factor mentioned above.

\subsection*{Acknowledgements}
I would like to thank Andris Ambainis for his suggestion to research on abilities concealed in the span programs.

This work has been supported by the European Social Fund within the project ``Support for Doctoral Studies at University of Latvia''.

\end{document}